\documentclass[prd,aps,preprintnumbers,nofootinbib,showpacs,superscriptaddress,floatfix,twocolumn]{revtex4}
\usepackage{epsfig}
\usepackage{graphicx}
\usepackage{amsmath}
\usepackage{amsfonts}
\usepackage{amssymb}

\allowdisplaybreaks

\begin{document}
\preprint{LYCEN 2008-03}
\title{A note on the determination of light quark masses}
\author{A. Deandrea}
\affiliation{Universit\'e de Lyon, IPNL, CNRS/IN2P3, 4 rue E. Fermi 69622 Villeurbanne cedex, France}
\author{A. Nehme}
\affiliation{Universit\'e de Lyon, IPNL, CNRS/IN2P3, 4 rue E. Fermi 69622 Villeurbanne cedex, France}
\affiliation{Lebanese University, Faculty of Sciences V, Nabatie, Lebanon}
\author{P. Talavera}
\affiliation{Universitat Polit\`ecnica de Catalunya,
Department de F{\'\i}sica i
Enginyeria Nuclear, Comte Urgell 187, E-08036 Barcelona, Spain}
\begin{abstract}
\noindent
We provide a model-independent determination of the quantity $B_0(m_d-m_u)$. Our approach rests only on chiral symmetry and data from the decay of the eta into three neutral pions. Since the low-energy prediction at next-to-leading order fails to reproduce the experimental results, we keep the strong interaction correction as an unknown parameter. As a first step, we relate this parameter to the quark mass difference using data from the Dalitz plot. A similar relation is obtained using data from the decay width. Combining both relations we obtain $B_0(m_d-m_u)=(4495\pm 440)\;\mathrm{MeV}^2$. The preceding value, combined with lattice determinations, leads to the values $ m_u(2\,\mathrm{GeV})=(2.9\pm 0.8)\,\mathrm{MeV} $ and $ m_d(2\,\mathrm{GeV})=(4.7\pm 0.8)\,\mathrm{MeV} $.
\end{abstract}
\maketitle

\section{Introduction \label{sec:I}}
One of the most reliable ways to extract predictions in the low-energy regime of strong interactions is the use of the chiral symmetry of QCD. This technique, suggested in \cite{Weinberg:1978kz}, was 
put on ground in \cite{GaLe1,GaLe2} and has been fully developed in chiral perturbation theory over the last 30 years. In the pion sector one finds strong evidence for the correctness and usefulness of chiral perturbation techniques. A further step consists on trying to extend this success into the strangeness  sector. The decay $ \eta\rightarrow\pi^0\pi^0\pi^0 $ is however not in good agreement with the experimental data, see for instance \cite{Holstein:2001bt}.  There can be several sources for this discrepancy: {\sl i)} The $\eta-\eta'$ mixing may play a major role in $\eta$ decays. However it was shown that within the chiral framework these effects are small for the decay $ \eta\rightarrow 3\pi$ \cite{Leutwyler:1996tz}. {\sl ii)} Final state interactions are significant in this channel \cite{Anisovich:1996tx,Kambor:1995yc}. {\sl iii)} The weight of the strange mass, suppressed by large-N$_c$ arguments at next-to-leading order, can re-emerge via unconstrained higher-order 
 counter-terms  \cite{Bijnens:2007pr}.
Most probably, as shown by the unitarization procedure of  \cite{Borasoy:2005du}, is some interplay between all these issues the cause of the disagreement. Our point of view evades a direct calculation in the strong sector and is instead based on the experimental data and chiral symmetry only. 

Chiral symmetry is explicitly broken by quark masses and electric charge. Quark masses are ad hoc parameters in the theory and their determination remains one of the most challenging tasks. On the experimental side, quark masses cannot be measured directly due to confinement. One has to relate them to observables which are sensitive to the variation of their values and try to extract the latter from the measurement of these observables. In the low-energy regime of the theory, quark masses come always multiplied by the quark condensate and one can \textit{only} determine quark mass ratios \cite{Gasser:1982ap,Leutwyler:1996qg}. Limiting ourselves to the light-quark sector including strangeness, these ratios are equal to one in the case of exact $ SU(3) $ flavor symmetry. Excluding strangeness, $ (m_u/m_d)=1 $ in the limit of exact $ SU(2) $ isospin symmetry. Therefore, deviation of the ratio $ m_u/m_d $ from unity can be detected in processes sensitive to isospin symmetry break
 ing. This is the case in the decay $ \eta\rightarrow 3\pi $ which is given mainly by the difference $ m_d-m_u $ \cite{Bell:1996mi,Baur:1995gc}. 

The paper is organized as follows. In Section \ref{sec:1} we give the analytic expression of the decay amplitude for the process $ \eta\rightarrow 3\pi^0 $ to one-loop order in Chiral Perturbation Theory (ChPT) including electromagnetic corrections. In Section \ref{sec:2} we derive a relation between the up and down quark mass difference and the strong part of the Dalitz plot parameter. We also give the ChPT prediction at one loop for the mass difference using data from the Dalitz plot. In Section \ref{sec:3} we use one-loop ChPT calculation for the decay width and the experimental value for the latter to predict the mass difference. Section \ref{sec:4} summarizes our method of extracting the mass difference from data only by relating the decay width to the Dalitz plot while keeping the strong interaction correction as a free parameter. In Section \ref{sec:5} we use our prediction for the mass difference to determine the size of the violation of Dashen's theorem and up and do
 wn quark masses.

\section{The decay amplitude \label{sec:1}}

Consider the process
\begin{equation}
\eta (p)\longrightarrow \pi^0(p_1)+\pi^0(p_2)+\pi^0(p_3)\,,
\end{equation}
with Mandelstam invariants
\begin{equation}
s\equiv (p-p_1)^2\,, \quad t\equiv (p-p_2)^2\,, \quad u\equiv (p-p_3)^2\,,
\end{equation}
satisfying
\begin{equation} \label{eq:neutral-mandelstam-relation} 
s+t+u=M_{\eta}^2+3M_{\pi^0}^2\equiv 3s_0\,.
\end{equation}
This is a $ 3 $-particle decay. Observables depend on $ 3\times 3 $ free variables. Conservation of $ 4 $-momentum reduces the number of free variables to $ 3\times 3-4 $. Since we are dealing with bosons, absence of spin implies that in the rest frame of the decaying particle, the orientation of the momentum configuration (the angles) is irrelevant and hence $ 3 $ variables are trivial. There remain $ 3\times 3-7 $ \textit{essential variables}, $ s $ and $ u $, say. 

\begin{figure}
\vskip 0.5truecm
\centering
\includegraphics[scale=0.7]{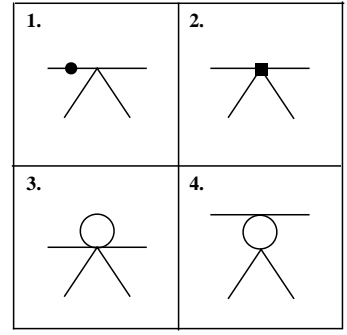} 
\caption{\label{fig:non-photonic} Feynman diagrams contributing to the decay amplitude at one loop. Diagram \textbf{1.} represents the wavefunction renormalization and the $\pi$--$\eta$ mixing. Diagram \textbf{2.} represents the counterterms. Diagram \textbf{3.} represents the tadpoles. Diagram \textbf{4.} represents the unitary corrections.}
\end{figure}

Let $ \Phi $ define the phase space integral
\begin{equation}
\Phi =\int\prod_{i=1}^3\dfrac{d^3p_i}{2E_i}\delta^{(4)}(p_1+p_2+p_3-p)\,.
\end{equation}
The phase space distribution is therefore a constant
\begin{equation} \label{eq:neutral-dalitz-density} 
\dfrac{d^2\Phi}{dsdu}=\dfrac{\pi^2}{4M_{\eta}^2}\,.
\end{equation}
It follows that the density of points in a Dalitz plot is proportional to the square of the matrix element. The physical region in the $ (s,u) $ plane is given by the boundary of the corresponding Dalitz plot
\begin{equation}
4M_{\pi^0}^2\leqslant s\leqslant (M_{\eta}-M_{\pi^0})^2\,, \quad u^-\leqslant u\leqslant u^+\,,
\end{equation}
where
\begin{equation}
u^{\pm}\,=\,\frac{1}{2}(3s_0-s)\pm \frac{1}{2}\sigma (s,M_{\pi^0}^2)\lambda^{1/2}(s,M_{\pi^0}^2,M_{\eta}^2)\,,
\end{equation}
and
\begin{eqnarray}
\sigma (x,m^2) &\equiv& \sqrt{1-\dfrac{4m^2}{x}}\,, \\
\lambda(x,m_1^2,m_2^2) &\equiv& \left[ x-(m_1-m_2)^2\right] \left[ x-(m_1+m_2)^2\right] \,.
\end{eqnarray}
Integration over $ u $ gives
\begin{equation}
\Phi =\dfrac{\pi^2}{4M_{\eta}^2}\int_{4M_{\pi^0}^2}^{(M_{\eta}-M_{\pi^0})^2}ds\,\sigma (s,M_{\pi^0}^2)\lambda^{1/2}(s,M_{\pi^0}^2,M_{\eta}^2)\,.
\end{equation}
The differential decay rate is defined as
\begin{equation}
d^2\Gamma =\dfrac{1}{2^6\pi^5M_{\eta}}\vert\mathcal{M}\vert^2d^2\Phi\,,
\end{equation}
and takes the following form
\begin{equation}
\dfrac{d^2\Gamma}{dsdu}=\dfrac{1}{2^8\pi^3M_{\eta}^3}\vert\mathcal{M}\vert^2\,,
\end{equation}
by (\ref{eq:neutral-dalitz-density}). The decay amplitude $ \mathcal{M} $ is defined as, 
\begin{equation}
\langle\pi^0\pi^0\pi^0|\eta\rangle\,\equiv\,i(2\pi)^4\delta^{(4)}(p_1+p_2+p_3-p)\mathcal{M}\,,
\end{equation}
and can be deduced from the set of Feynman diagrams depicted in Fig. \ref{fig:non-photonic}. 
In order to evaluate these diagrams, one has to expand the leading (next-to-leading) order chiral Lagrangian to the sixth (fourth) order in the meson field and to first (zeroth) order in the photon field. Let $q^2$ and $m_q$ represent generic energy and generic light quark mass, respectively. Chiral power counting of the diagrams shows that the amplitude counts like $q^2$ or $m_q$ at tree level and $q^4$, $m_q^2$ or $m_qq^2$ at one-loop level in the isospin limit and in the classic standard scheme, that is, $q^2\sim m_q$. In the isospin breaking case, additional terms arise from the two sources of isospin breaking, namely, the electric charge and the up and down quark mass difference. From the electromagnetic mass of the pion,
\begin{equation}
\Delta_{\pi}\,\equiv\,M_{\pi^{\pm}}^2-M_{\pi^0}^2\,=\,\mathcal{O}(e^2)\,,
\end{equation} 
we see that $e^2$ counts as $m_q$. It follows that the additional terms count like $ e^2 $ or $ m_d-m_u $ at tree level, and, $ q^2(m_d-m_u) $, $ m_q(m_d-m_u) $, $ (m_d-m_u)^2 $, $ e^2(m_d-m_u) $, $ e^2q^2 $, $ e^2m_q $, or $e^4$ at one-loop level. By naive dimensional analysis, the $e^4$ and $ (m_d-m_u)^2 $ terms are suppressed with respect to the others and henceforth will be neglected.

The decay amplitude can be written by symmetry considerations as
\begin{equation} \label{eq:neytral-amplitude-symmetry} 
\mathcal{M}(s,t,u)= M(s)+M(t)+M(u)\,.
\end{equation}
The $ s $-channel amplitude can be casted in the following general form
\begin{eqnarray} \label{eq:amplitude} 
M(s)\,=\,-\dfrac{B_0(m_d-m_u)}{3\sqrt{3}F_{\pi}^2}\left\lbrace 1+\delta_{\mathrm{str}}(s)+\delta_{\mathrm{em}}(s)\right\rbrace +\tilde{\delta}_{\mathrm{em}}(s)\,,
\end{eqnarray}
where $ B_0 $ is an order parameter for chiral symmetry which is related to the quark condensate,
\begin{equation} \label{eq:condensate} 
B_0\,=\,-\displaystyle\lim_{\substack{m_q\rightarrow 0\\e\rightarrow 0}}\dfrac{\langle\bar{q}q\rangle}{F_{\pi}^2}\,,
\end{equation}
and $F_{\pi}$ is the pion decay constant.

All three corrections in (\ref{eq:amplitude}) are separately ultraviolet finite. Both of them, $\delta_{\rm str}$ and $\tilde{\delta}_{\rm em}$, have been already considered in the literature,
\cite{Gasser:1984pr,Baur:1995gc} respectively. In addition to these we have included a small, with respect to the previous ones, term proportional to $(m_d-m_u) e^2$. For completeness, we give the analytic expressions for the corrections in the Appendix.

\section{The Dalitz plot\label{sec:2}}

The Dalitz plot distribution is parameterized as
\begin{equation} \label{eq:parametrization} 
\vert\mathcal{M}(s,t,u)\vert^2\,=\,\vert\mathcal{M}(s_0,s_0,s_0)\vert^2\left\lbrace 1+2\,\alpha \,z\right\rbrace \,,
\end{equation}
with $ s_0 $ the center of the Dalitz plot and $ z $ a dimensionless parameter defined in terms of pion energies $ E_i $ as
\begin{equation}
z=\frac{2}{3}\sum_{i=1}^3\left( \dfrac{3E_i-M_{\eta}}{M_{\eta}-3M_{\pi^0}}\right) ^2\,,
\end{equation}
and in terms of Mandelstam variables like
\begin{equation} \label{eq:neutral-center-dalitz-slope} 
z=\dfrac{3}{2}\dfrac{1}{M_{\eta}^2(M_{\eta}-3M_{\pi^0})^2}\left\lbrace (s-s_0)^2+(t-s_0)^2+(u-s_0)^2\right\rbrace \,.
\end{equation}
The experimental value of $ \alpha $ as quoted in the PDG \cite{Yao:2006px} is
\begin{equation} \label{eq:alpha-experimental} 
\alpha_{\mathrm{exp}}\,=\,-0.031\pm 0.004\,.
\end{equation} 
In order to conform to the experimental analysis, we expand the square of the amplitude around the center of the Dalitz plot with the use of
\begin{eqnarray}
&&\mathcal{M}(s,t,u) = \sum_{j=0}^{\infty}\left\lbrace \frac{1}{j!}\left[ (s-s_0)\dfrac{\partial}{\partial s'} \right. \right. \nonumber\\
&& \left. \left. +(t-s_0)\frac{\partial}{\partial t'}+(u-s_0)\frac{\partial}{\partial u'}\right] ^j\mathcal{M}(s',t',u')\right\rbrace _{s'=t'=u'=s_0}\,.
\end{eqnarray}
We then match with the parametrization (\ref{eq:parametrization}), obtaining the scale-independent 
expression
\begin{equation} \label{eq:formula} 
B_0(m_d-m_u)\,=\,\dfrac{\tilde{\alpha}_{\mathrm{em}}}{\alpha_{\mathrm{exp}}-\alpha_{\mathrm{str}}-\alpha_{\mathrm{em}}}\,\Delta_{\pi}\,.
\end{equation}
Even if we have obtained (\ref{eq:formula} ) at next-to-leading order is straightforward to convince oneself that its 
explicit form is maintained at all orders in the strong interaction and at first order in the electromagnetic one. Note that the $ \alpha $'s are given by the second derivative of the corresponding $ \delta $'s, up to normalization factors, evaluated at the center of the Dalitz,
\begin{eqnarray}
\label{als}
\alpha_{\mathrm{str}} &=& \frac{1}{9}M_{\eta}^2(M_{\eta}-3M_{\pi^0})^2\,\mathrm{Re}\,\delta_{\mathrm{str}}''(s_0)\,, \\
\alpha_{\mathrm{em}} &=& \frac{1}{9}M_{\eta}^2(M_{\eta}-3M_{\pi^0})^2\,\mathrm{Re}\,\delta_{\mathrm{em}}''(s_0)\,, \\
\Delta_{\pi}\tilde{\alpha}_{\mathrm{em}} &=& -\dfrac{F_{\pi}^2}{\sqrt{3}}M_{\eta}^2(M_{\eta}-3M_{\pi^0})^2\,\mathrm{Re}\,\tilde{\delta}_{\mathrm{em}}''(s_0)\,.
\end{eqnarray}
Although (\ref{eq:formula}) expresses the up and down quark mass difference in terms of the electric charge $e$, it does \textit{not} provide the electromagnetic part of that difference. In fact, taking the limit of vanishing $e$, both numerator and denominator in (\ref{eq:formula}) tend to zero and one \textit{cannot} claim that $ e\rightarrow 0$ implies $m_u=m_d$. On the other hand, expression 
(\ref{eq:formula}) relates up and down quark mass difference to the electric charge through an observable, namely, 
$\alpha_{\mathrm{exp}}$.

As a first step, we obtain the analytic expressions for the $\alpha$'s to first order in the chiral expansion. For compactness, we use the Gell-Mann--Okubo mass formula, $ 4M_K^2=3M_{\eta}^2+M_{\pi}^2 $, and obtain the reduced expressions for the $\alpha$'s given in the Appendix.

The main feature of these expressions is the independence on any low-energy constant. This can already be seen from the fact that the coefficients of the latter in the $\delta$'s are polynomials in meson masses and do not depend on Mandelstam invariants. Taking the second derivative with respect to these invariants gives a null contribution to the low-energy constants. The numerical values, at next-to-leading order, of the $\alpha$'s are found to be 
\begin{equation}
\label{one}
\alpha_{\mathrm{str}}^{(1)}\,=\,+0.0179\,,
\end{equation}
for the strong piece, and
\begin{equation}
\alpha_{\mathrm{em}}^{(1)}\,=\,-0.0011\,, \quad \tilde{\alpha}_{\mathrm{em}}^{(1)}\,=\,0.0025\,,
\end{equation}
for the electromagnetic ones. 
Notice that (\ref{eq:alpha-experimental}) is roughly a factor two bigger and of opposite sign of (\ref{one}).
Replacing the $\alpha$'s by their values in (\ref{eq:formula}), we obtain at one-loop order, 
\begin{equation} \label{eq:difference-dalitz} 
B_0(m_d-m_u)^{(1)}\,=\,-(66\pm 5)\;\mathrm{MeV}^2\,.
\end{equation}
As $B_0$ must be positive definite the previous expression leads to the conclusion that the up quark is heavier than the down quark. The situation does not improve when including higher-order corrections or rescattering effects in the final state (see Table 5 of \cite{Bijnens:2007pr}). Moreover a $10\%$  correction to the value of (\ref{one}) roughly translates in a $4\%$ correction to (\ref{eq:difference-dalitz}). A close inspection of (\ref{eq:formula}) reveals that, disregarding $\alpha_{\rm em}$, this inconsistency will always show up if
$\alpha_{\rm str}$ is not bigger, in absolute value, than $\alpha_{\rm exp}$. Notice that this is against {\sl all} the findings in the literature tackling so far the strong sector. We shall attempt in the next Sections a different approach using a parametrization of available experimental data.

\section{The decay width \label{sec:3}}

As in the preceding section, we can repeat a similar procedure for the total decay rate. The decay rate takes the following form
\begin{eqnarray} \label{eq:width} 
\Gamma &=& B_0^2(m_d-m_u)^2\left( \gamma_{\mathrm{tree}}+\gamma_{\mathrm{str}}+\gamma_{\mathrm{em}}\right) \nonumber\\
&& +B_0(m_d-m_u)\,\tilde{\gamma}_{\mathrm{em}}\,,
\end{eqnarray}
where the $\gamma$'s are obtained, up to a normalization factor, by integrating the $\delta$'s over the allowed kinematical region,
\begin{eqnarray}
\gamma_{\mathrm{tree}} &=& \dfrac{1}{3F_{\pi}^4}\hat{\mathcal{F}}\,1\,, \\
\gamma_{\mathrm{str}} &=& \dfrac{2}{3F_{\pi}^4}\hat{\mathcal{F}}\,\mathrm{Re}\,\delta_{\mathrm{str}}(s)\,, \\
\gamma_{\mathrm{em}} &=& \dfrac{2}{3F_{\pi}^4}\hat{\mathcal{F}}\,\mathrm{Re}\,\delta_{\mathrm{em}}(s)\,, \\
\tilde{\gamma}_{\mathrm{em}} &=& -\dfrac{6}{\sqrt{3}F_{\pi}^2}\hat{\mathcal{F}}\,\mathrm{Re}\,\tilde{\delta}_{\mathrm{em}}(s)\,. \\
\end{eqnarray}
The operator $ \hat{\mathcal{F}} $ is defined as
\begin{widetext}
\begin{equation}
\hat{\mathcal{F}}f(s) = \dfrac{1}{2^8\pi^3M_{\eta}^3}\int_{4M_{\pi^0}^2}^{(M_{\eta}-M_{\pi^0})^2}ds\sigma (s,M_{\pi^0}^2)\lambda^{1/2}(s,M_{\pi^0}^2,M_{\eta}^2)f(s)\,.
\end{equation}
\end{widetext}
The experimental value of the rate as obtained from the PDG is
\begin{equation} \label{eq:gamma-experimental} 
\Gamma_{\mathrm{exp}}\,=\,(421\pm 22)\;\mathrm{eV}\,.
\end{equation}
Using for the amplitude (\ref{eq:amplitude}) the Born approximation we obtain, 
\begin{equation}
\gamma_{\mathrm{tree}}\,=\,22.31\times 10^{-12}\;\mathrm{MeV}^{-3}\,, 
\end{equation}
while using the next-to-leading expression the contributions are split as
\begin{equation}
\label{eq:one}
\gamma_{\mathrm{str}}^{(1)}\,=\,(35.70\pm 14.14)\times 10^{-12}\;\mathrm{MeV}^{-3}\,,
\end{equation}
in the strong sector and
\begin{eqnarray}
\gamma_{\mathrm{em}}^{(1)} &=& -(0.02\pm 0.34)\times 10^{-12}\;\mathrm{MeV}^{-3}\,, \\
\tilde{\gamma}_{\mathrm{em}}^{(1)} &=& -(2.50\pm 2.44)\times 10^{-9}\;\mathrm{MeV}^{-1}\,,
\label{twoo}
\end{eqnarray}
in the electromagnetic one. Replacing the $ \gamma $'s by their values in (\ref{eq:width}), solving for the quark mass difference and discarding the negative root we obtain from $ \Gamma_{\mathrm{exp}} $ the next-to-leading order value
\begin{equation} \label{eq:difference-width} 
B_0(m_d-m_u)^{(1)}\,=\,(2717\pm 342)\;\mathrm{MeV}^2\,,
\end{equation}
which can be afflicted with sizable and unknown higher order contributions.  

Hitherto we have followed two different paths to obtain $B_0(m_d-m_u)$, leading to incompatible results:
{\sl i )} First we calculated the $ \delta $'s in (\ref{eq:amplitude}) at one-loop in Chiral Perturbation Theory. After we evaluated their second derivative at the center of the Dalitz in order to obtain the $ \alpha $'s (\ref{als}). We matched the latter with the data coming from the Dalitz and obtained for the quark mass difference the puzzling value (\ref{eq:difference-dalitz}).
{\sl ii)}  As a second approach we integrated the very same expressions for the $\delta $'s over the physical region in order to obtain the $ \gamma $'s, (\ref{eq:one}--\ref{twoo}). We matched the latter with the data coming from the decay width and obtained for the difference the value (\ref{eq:difference-width}). 
Both results, (\ref{eq:difference-dalitz}) and (\ref{eq:difference-width}) are in disagreement.
This observation is a sufficient motivation to seek for a determination of these quantities that is as model independent as possible. We shall therefore ignore the theoretical determinations concerning the strong interaction and provide a determination merely based on chiral symmetry and on a combination of data from the Dalitz plot and the decay rate. 

\section{Relating the Dalitz to the width \label{sec:4}}

The decay rate can be obtained by applying two equivalent methods: {\sl i)} Either integrating the full amplitude, expression (\ref{eq:amplitude}). {\sl ii)} Or one can instead integrate the parametrization for the amplitude  (\ref{eq:parametrization}). By working out explicitly both approaches and matching them, one can relate the quark mass difference to the Dalitz parameter in the strong sector, $ \alpha_{\mathrm{str}} $ without resorting to any definite value for $\delta_{\mathrm{str}}$. In doing so, we avoid attempting to dig out the how to obtain the piece $\delta_{\mathrm{str}}$ from first principles and we leave it as a free parameter.
This is justified by comparing the relative size of the perturbative values for the $\delta$'s at next-to-leading order and the expectation of strong rescattering effects at higher order. 

As a first step we find a relation between $ \mathrm{Re}\,\delta_{\mathrm{str}}(s_0) $ and $ B_0(m_d-m_u) $. For that purpose we integrate (\ref{eq:parametrization}) over the whole Dalitz to obtain the decay rate. Using for $ \alpha $ and $ \Gamma $ their experimental values given respectively by (\ref{eq:alpha-experimental}) and (\ref{eq:gamma-experimental}) we determine the distribution at the center
\begin{equation} \label{eq:distribution-center} 
\vert\mathcal{M}(s_0,s_0,s_0)\vert^2\,=\,(88.67\pm 4.57)\times 10^{-3}\,.
\end{equation}
>From the previous value and the expression of the distribution at the center, 
\begin{widetext}
\begin{eqnarray}
\vert\mathcal{M}(s_0,s_0,s_0)\vert^2 &=& \dfrac{1}{3F_{\pi}^4}B_0^2(m_d-m_u)^2\,\{1+2\,\mathrm{Re}\,\delta_{\mathrm{str}}(s_0)+2\,\mathrm{Re}\,\delta_{\mathrm{em}}(s_0)\,\}
-\dfrac{6}{\sqrt{3}F_{\pi}^2}B_0(m_d-m_u)\,\mathrm{Re}\,\tilde{\delta}_{\mathrm{em}}(s_0)\,,
\end{eqnarray}
\end{widetext}
derived from (\ref{eq:amplitude}), we write $ \mathrm{Re}\,\delta_{\mathrm{str}}(s_0) $ in terms of $ B_0(m_d-m_u) $. We shall make use of this relation latter in order to pull out $ \mathrm{Re}\,\delta_{\mathrm{str}}(s_0) $ in favor of $ B_0(m_d-m_u) $.

For the last step we start from the definition of $ \gamma_{\mathrm{str}} $ in terms of $ \delta_{\mathrm{str}} $ in (\ref{eq:width}). Expanding the latter around the center of the Dalitz and integrating we obtain a relation between $ \gamma_{\mathrm{str}} $, $ \alpha_{\mathrm{str}} $ and $\mathrm{Re}\,\delta_{\mathrm{str}}(s_0) $,
\begin{equation}
\gamma_{\mathrm{str}}\,=\,2\gamma_{\mathrm{tree}}\,\mathrm{Re}\,\delta_{\mathrm{str}}(s_0)+\dfrac{1}{3F_{\pi}^4}\dfrac{9\alpha_{\mathrm{str}}}{M_{\eta}^2(M_{\eta}-3M_{\pi^0})^2}\hat{\mathcal{F}}\,(s-s_0)^2\,.
\end{equation} 
Equating now the two expressions for $ \gamma_{\mathrm{str}} $ we obtain a second (independent) relation between $ \mathrm{Re}\,\delta_{\mathrm{str}}(s_0) $ and $ B_0(m_d-m_u) $. Solving for $ B_0(m_d-m_u) $ leads to the final expression
\begin{equation}
 \label{eq:relation} 
(\alpha_{\mathrm{str}}+a_{\mathrm{em}})B_0^2(m_d-m_u)^2 +\tilde{a}_{\mathrm{em}}B_0(m_d-m_u)+a_{\mathrm{exp}}=0\,,
\end{equation}
with
$$
a_{\mathrm{em}} = \frac{2}{9}M_{\eta}^2(M_{\eta}-3M_{\pi^0})^2 
\dfrac{\hat{\mathcal{F}}\,\mathrm{Re}\left[ \delta_{\mathrm{em}}(s)-\delta_{\mathrm{em}}(s_0)\right]}{\hat{\mathcal{F}}(s-s_0)^2}\,,  
$$
$$
\tilde{a}_{\mathrm{em}} = -\frac{2F_{\pi}^2}{\sqrt{3}}M_{\eta}^2(M_{\eta}-3M_{\pi^0})^2
\dfrac{\hat{\mathcal{F}}\,\mathrm{Re}\left[ \tilde{\delta}_{\mathrm{em}}(s)-\tilde{\delta}_{\mathrm{em}}(s_0)\right] }{\hat{\mathcal{F}}(s-s_0)^2}\,, 
$$
\begin{equation}
a_{\mathrm{exp}} = -\dfrac{3F_{\pi}^4M_{\eta}^2(M_{\eta}-3M_{\pi^0})^2\alpha_{\mathrm{exp}}\Gamma_{\mathrm{exp}}}{9\alpha_{\mathrm{exp}}\hat{\mathcal{F}}(s-s_0)^2+M_{\eta}^2(M_{\eta}-3M_{\pi^0})^2\hat{\mathcal{F}}1}\,. 
\end{equation}
At next-to-leading order in the chiral counting 
\begin{eqnarray}
a_{\mathrm{em}}^{(1)} &=& 145.374\times 10^{-6}\,, \\
\tilde{a}_{\mathrm{em}}^{(1)} &=& 2.905\;\mathrm{MeV}^2\,, \\
a_{\mathrm{exp}} &=& (0.601\pm 0.085)\times 10^6\;\mathrm{MeV}^4\,.
\end{eqnarray}
Note that the preceding coefficients are low-energy-constant independent. This is due to the fact that, in their analytic expression, we integrate the real part of the $\delta$'s subtracted at the center of the Dalitz, that is, $ \delta (s)-\delta (s_0)$. Since the coefficients of the low-energy constants in the $\delta$'s have no kinematical dependence, the subtraction allows to get rid of the low-energy constants. In other terms, the origin of the errors in the $a$'s, which is also the case for the $\alpha$'s, is purely experimental.

We have now the two relations (\ref{eq:formula}) and (\ref{eq:relation}) with the two unknowns, $ B_0(m_d-m_u) $ and $ \alpha_{\mathrm{str}}$. The first relation comes from the Dalitz alone. The second relation is a combination of data from both the Dalitz and the width. The solution of the system, constraining the down quark to be heavier than the up quark, gives the following numbers
\begin{eqnarray}
\alpha_{\mathrm{str}} &=& -(0.0305\pm 0.004)\,, \label{eq:alpha} \\
B_0(m_d-m_u) &=& (4495\pm 440)\;\mathrm{MeV}^2\,. \label{eq:difference} 
\end{eqnarray}
If instead we use the recent data from KLOE \cite{Ambrosino:2007wi} we find 
$\alpha_{\mathrm{str}}=-(0.0265\pm 0.0100)$ for the strong interaction Dalitz parameter and $B_0(m_d-m_u)=(4832\pm 999)\;\mathrm{MeV}^2$ for the quark mass difference. 

\section{Survey of applications \label{sec:5}}

\subsection{Violation of Dashen's theorem}

We consider the square mass difference
$\Delta_K-\Delta_{\pi}$,
which can be split into two pieces
\begin{equation} \label{eq:splitting} 
\Delta_K-\Delta_{\pi}\,=\,(\Delta_K)_{\mathrm{str}}+(\Delta_K-\Delta_{\pi})_{\mathrm{em}}\,,
\end{equation}
where $\Delta_P\equiv M_{P^{\pm}}^2-M_{P^0}^2$.
We have used the fact that the pion mass difference is essentially of electromagnetic origin \cite{Das:1967it}. 
The electromagnetic term in (\ref{eq:splitting}) vanishes at leading chiral order by virtue of Dashen's theorem \cite{Dashen:1969eg} but is subject to corrections from higher orders. These corrections are commonly known as the violation of Dashen's theorem. In order to estimate the size of the violation we shall introduce the parameter $ \kappa $ defined by
\begin{equation}
\left( M_{K^{\pm}}^2-M_{K^0}^2\right) _{\mathrm{em}}\,\equiv \,\kappa\left( M_{\pi^{\pm}}^2-M_{\pi^0}^2\right) _{\mathrm{em}}\,,
\end{equation}
and rewrite the difference (\ref{eq:splitting}) at next-to-leading order in the following ultraviolet finite form
\begin{equation}\label{deltm}
(\Delta_K-\Delta_{\pi})^{(1)} = -B_0(m_d-m_u)
\left\lbrace 1+\Delta_{\mathrm{str}}+\Delta_{\mathrm{em}}\right\rbrace +\tilde{\Delta}_{\mathrm{em}}\,.
\end{equation}
The origin of the different terms in (\ref{deltm}) runs parallel to those of (\ref{eq:amplitude}): $\Delta_{\mathrm{str}}$ and $\tilde{\Delta}_{\mathrm{em}}$ where obtained already in \cite{GaLe2,Urech:1994hd}, while the sub-leading $\Delta_{\mathrm{em}}$ term is new. 
Numerically its contribution turns to be around $8\%-20\%$ corrections to the $\tilde{\Delta}_{\rm em}$ value depending on the strong low-energy constant set we use. At next-to-leading order their numerical values are given by
\begin{eqnarray}
\Delta_{\mathrm{str}} &=& 0.277\pm 0.340\,, \\ \label{firstdel}
\Delta_{\mathrm{em}} &=& -(0.0014\pm 0.0055)\,, \\
\tilde{\Delta}_{\mathrm{em}} &=& -(194\pm 1517)\;\mathrm{MeV}^2\,,
\label{delemt}
\end{eqnarray}
but these are subject to the uncertainty in the low-energy constants, as is indeed mainly reflected in (\ref{delemt}).
Comparing (\ref{deltm}) with (\ref{eq:splitting}) and using the estimates (\ref{firstdel}--\ref{delemt})
we obtain,
\begin{equation}
\kappa^{(1)} = 0.85\pm 1.20\,.
\end{equation}
The central value can be considered moderate in front of other estimates \cite{Bijnens:1996kk} although no solid conclusion can be obtained in view of the errors. 
An alternative, and more error free, approach consists on replacing in (\ref{eq:splitting}) the total difference by physical masses, the strong piece by its one-loop value and deduce the value of the electromagnetic piece and, \textit{a fortiori}, that of $ \kappa $. We find
\begin{equation}
\label{dashdos}
\kappa\,=\,1.4\pm 1.3\,.
\end{equation}
Notice that this value of $ \kappa $ contains, beside the violation of Dashen's theorem, the effect of higher-order 
strong corrections which can be sizeable. 

\subsection{Light quark masses}

The parameter $ B_0 $ is defined by Eq.(\ref{eq:condensate}). We take for the pion decay constant  and for the quark condensate in the chiral limit  the unquenched lattice determination \cite{Bernard:2007ps}, $ F_0(2\,\mathrm{GeV})= (76\pm 3)\,\mathrm{MeV} $ and $ \langle\bar{q}q\rangle_0(2\,\mathrm{GeV})=-(242\pm 9)^3\,\mathrm{MeV}^3 $, respectively. This leads to the value $ B_0(2\,\mathrm{GeV})=(2454\pm 194)\,\mathrm{MeV} $ where the main source of error comes from $F_0$. Together with Eq. (\ref{eq:difference}), the preceding estimation leads to
\begin{equation}
(m_d-m_u)(2\,\mathrm{GeV})\,=\,(1.8\pm 0.2)\,\mathrm{MeV}\,.
\end{equation}
Now we use for $ m_d+m_u $ the lattice determination quoted in \cite{Yao:2006px}, namely, $ (m_d+m_u)(2\,\mathrm{GeV})=(7.6\pm 1.6)\,\mathrm{MeV} $, and obtain
\begin{eqnarray} \label{eq:quark-masses} 
m_u(2\,\mathrm{GeV})&=&(2.9\pm 0.8)\,\mathrm{MeV}\\ 
m_d(2\,\mathrm{GeV})&=&(4.7\pm 0.8)\,\mathrm{MeV}\,.
\end{eqnarray}
Both results are, within errors, in agreement with the more recent lattice results either using domain wall \cite{Allton:2008pn} \cite{Blum:2007cy}, improved Wilson \cite{Ukita:2007cu}, twisted mass \cite{Blossier:2007vv} or staggered fermions \cite{Bernard:2007ps}.

\section{Conclusion \label{sec:C}}
In this letter we have used chiral symmetry and data to obtain a consistent parametrization of 
the $\eta \to 3 \pi^0$ decay. From the theoretical side we have pointed out that the calculation of the 
strong interaction effects are the most probable source of disagreement with data. We therefore take 
this contribution as a free parameter rather than attempting to evaluate it. The electromagnetic contributions
are instead small with respect to the strong ones and rather well understood. We incorporate them 
using Chiral Perturbation Theory. Combining the Dalitz plot parametrization with data we obtained a relation between the strong interaction parameter $\alpha_{\mathrm{str}}$ and the quantity $B_0(m_d-m_u)$. 
Following a similar procedure for the width we obtained a second relation between the same quantities. Solving the system for both unknowns we obtained $\alpha_{\mathrm{str}}= -(0.0305\pm 0.004)$ and
$B_0(m_d-m_u)=(4495\pm 440)\;\mathrm{MeV}^2$. As a first application, we estimated roughly the size of the 
violation of Dashen's theorem (\ref{dashdos}). As a second application, we estimated the values of up and down quark masses (\ref{eq:quark-masses}).

\section{Appendix}
We give here for completeness the expressions for the $\delta$s and the $\alpha$s appearing in the text 
\begin{widetext}
\begin{eqnarray}
\delta_{\mathrm{str}} &=& -\dfrac{1}{240F_{\pi}^2}\dfrac{1}{M_K^2}\left\lbrace \dfrac{1}{16\pi^2}\bigg [135 s^2
-15s(18M_K^2+M_{\pi}^2+3M_{\eta}^2)+2M_K^2(16M_K^2-49M_{\pi}^2+13M_{\eta}^2)\bigg ] \right.\nonumber\\
&& +320M_K^2\bigg [ 6(2M_K^2+M_{\pi}^2)L_4+(4M_K^2+11M_{\pi}^2+3M_{\eta}^2)L_5
-12(2M_K^2+M_{\pi}^2)L_6+48(M_K^2-4M_{\pi}^2)L_7-96M_{\pi}^2L_8\bigg ] \nonumber\\
&& +\left( \dfrac{4M_K^2}{M_K^2-M_{\pi}^2}\right) (150s-62M_K^2-27M_{\pi}^2-26M_{\eta}^2)A(M_{\pi}^2)
-\left( \dfrac{20M_K^2}{M_K^2-M_{\pi}^2}\right) (2M_K^2+M_{\pi}^2)A(M_{\eta}^2) \nonumber\\
&& +\left( \dfrac{2}{M_K^2-M_{\pi}^2}\right) \bigg [ (13M_{\eta}^2-75s)(3M_K^2+M_{\pi}^2)
+168M_K^4+49M_{\pi}^2M_K^2-49M_{\pi}^4\bigg ] A(M_K^2) \nonumber\\
&& -\left( \dfrac{60M_K^2}{M_K^2-M_{\pi}^2}\right) \bigg [ (9s-22M_{\pi}^2-3M_{\eta}^2+4M_K^2)s
+M_{\pi}^2(2M_K^2+7M_{\pi}^2+3M_{\eta}^2)\bigg ] B(s,M_{\pi}^2,M_{\pi}^2) \nonumber\\
&& +120M_{\pi}^2M_K^2B(s,M_{\eta}^2,M_{\eta}^2)-240M_{\pi}^2M_K^2B(s,M_{\pi}^2,M_{\eta}^2)
+\left( \dfrac{60M_K^2}{M_K^2-M_{\pi}^2}\right) \bigg [ (9s-14M_K^2-4M_{\pi}^2-3M_{\eta}^2)s \nonumber\\
&& +M_K^2(8M_K^2+M_{\pi}^2+3M_{\eta}^2)\bigg ] B(s,M_K^2,M_K^2)
\left. +15s^2\bigg [ 9s-(8M_K^2+M_{\pi}^2+3M_{\eta}^2)\bigg ] B'(s,M_K^2,M_K^2)\right\rbrace \,, \\
\delta_{\mathrm{em}} &=& -\dfrac{2e^2}{9}\bigg\{ 12(K_1+K_2)-12(K_7+K_8)
-6(2K_3-K_4)+8(K_5+K_6)-10(K_9+K_{10})\bigg\} \nonumber\\
&& +\dfrac{e^2}{3}\left( \dfrac{M_{\pi}^2}{M_K^2-M_{\pi}^2}\right) \bigg\{ 3(2K_3-K_4)-2(K_5+K_6)\bigg\}
-\dfrac{1}{480F_{\pi}^2}\dfrac{1}{M_{\pi}^2M_K^2}\left( \dfrac{\Delta_{\pi}}{M_K^2}\right) \times \nonumber\\
&& \left\lbrace \dfrac{1}{16\pi^2}\dfrac{1}{M_K^2-M_{\pi}^2}\bigg [ 135(8M_K^4-5M_{\pi}^2M_K^2+M_{\pi}^4)s^2 \right. 
-5(486M_K^4M_{\pi}^2-57M_{\pi}^4M_K^2+3M_{\pi}^6+9M_{\pi}^4M_{\eta}^2 \nonumber\\
&& -45M_{\pi}^2M_K^2M_{\eta}^2-96M_K^6+72M_K^4M_{\eta}^2)s
-2M_{\pi}^2M_K^2(520M_K^4-909M_{\pi}^2M_K^2 \nonumber\\
&& -49M_{\pi}^4+13M_{\pi}^2M_{\eta}^2-155M_K^2M_{\eta}^2)\bigg ]
+\left( \dfrac{4M_K^4}{M_K^2-M_{\pi}^2}\right) (300s-64M_K^2-104M_{\pi}^2-52M_{\eta}^2)A(M_{\pi}^2) \nonumber\\
&& -\left( \dfrac{4M_{\pi}^2M_K^2}{M_K^2-M_{\pi}^2}\right) (150s-92M_K^2+8M_{\pi}^2-26M_{\eta}^2)A(M_K^2)
+\left( \dfrac{60sM_K^4}{M_K^2-M_{\pi}^2}\right) \bigg [ 18s^2+2(4M_K^2-3M_{\eta}^2-22M_{\pi}^2)s \nonumber\\
&& -M_{\pi}^2(8M_K^2-26M_{\pi}^2-6M_{\eta}^2)\bigg ] B'(s,M_{\pi}^2,M_{\pi}^2)
-\left( \dfrac{30sM_{\pi}^2}{M_K^2-M_{\pi}^2}\right) \bigg [ 9(3M_K^2-M_{\pi}^2)s^2 \nonumber\\
&& -(36M_K^4+M_{\pi}^2M_K^2-M_{\pi}^4+9M_K^2M_{\eta}^2-3M_{\pi}^2M_{\eta}^2)s
+2M_K^4(8M_K^2+M_{\pi}^2+3M_{\eta}^2)\bigg ] B'(s,M_K^2,M_K^2) \nonumber\\
&& -15s^3M_{\pi}^2(9s-8M_K^2-M_{\pi}^2-3M_{\eta}^2)B''(s,M_K^2,M_K^2)\bigg\}\,, \\
\tilde{\delta}_{\mathrm{em}} &=& \dfrac{2e^2M_{\pi}^2}{9\sqrt{3}F_{\pi}^2}\bigg\{3(2K_3-K_4)-2(K_5+K_6)+2(K_9+K_{10})\bigg\}
+\dfrac{1}{720\sqrt{3}F_0^4}\left( \dfrac{2Z_0e^2F_0^2}{M_K^2}\right) \left\lbrace \dfrac{1}{16\pi^2}\bigg [ -135s^2 \right. \nonumber\\
&& +15(18M_K^2+M_{\pi}^2+3M_{\eta}^2)s-M_K^2(32M_K^2-98M_{\pi}^2+26M_{\eta}^2)\bigg ] \nonumber\\
&& +2(-75s+16M_K^2-49M_{\pi}^2+13M_{\eta}^2)A(M_K^2)
+15s^2(-9s+8M_K^2+M_{\pi}^2+3M_{\eta}^2)B'(s,M_K^2,M_K^2)\bigg\}\,.
\end{eqnarray}

\begin{eqnarray}
\alpha_{\mathrm{str}}^{(1)} &=& \dfrac{1}{108F_{\pi}^2}\dfrac{M_{\eta}^2(M_{\eta}-3M_{\pi})^2}{(M_{\eta}^2-M_{\pi}^2)(M_{\pi}^2+3M_{\eta}^2)}\times
\mathrm{Re}\,\{ -54(4\pi )^{-2}(M_{\eta}^2-M_{\pi}^2) \nonumber\\
&& -72(M_{\pi}^2+3M_{\eta}^2)B(s_0,M_K^2,M_K^2)
+12(M_{\pi}^2-5M_{\eta}^2)(5M_{\pi}^2-3M_{\eta}^2)B'(s_0,M_K^2,M_K^2) \nonumber\\
&& -(M_{\eta}^2-M_{\pi}^2)(117M_{\pi}^4-26M_{\pi}^2M_{\eta}^2+21M_{\eta}^4)B''(s_0,M_K^2,M_K^2)
+2(M_{\eta}^2-M_{\pi}^2)^2(3M_{\pi}^2+M_{\eta}^2)^2B^{(3)}(s_0,M_K^2,M_K^2) \nonumber\\
&& +72(M_{\pi}^2+3M_{\eta}^2)B(s_0,M_{\pi}^2,M_{\pi}^2)
-24(M_{\pi}^2+3M_{\eta}^2)(M_{\pi}^2-2M_{\eta}^2)B'(s_0,M_{\pi}^2,M_{\pi}^2) \nonumber\\
&& +2(M_{\eta}^2-M_{\pi}^2)(M_{\pi}^2+3M_{\eta}^2)(9M_{\pi}^2+2M_{\eta}^2)B''(s_0,M_{\pi}^2,M_{\pi}^2)
-6M_{\pi}^2(M_{\eta}^2-M_{\pi}^2)(M_{\pi}^2+3M_{\eta}^2)B''(s_0,M_{\eta}^2,M_{\eta}^2) \nonumber\\
&& +12M_{\pi}^2(M_{\eta}^2-M_{\pi}^2)(M_{\pi}^2+3M_{\eta}^2)B''(s_0,M_{\pi}^2,M_{\eta}^2)\,\}\,, \\
\alpha_{\mathrm{em}}^{(1)} &=& \dfrac{1}{162F_{\pi}^2}\dfrac{\Delta_{\pi}}{M_{\pi}^2}\dfrac{M_{\eta}^2(M_{\eta}-3M_{\pi})^2}{(M_{\eta}^2-M_{\pi}^2)(M_{\pi}^2+3M_{\eta}^2)^2}\times
\mathrm{Re}\,\{ -54(4\pi )^{-2}(M_{\pi}^4-3M_{\pi}^2M_{\eta}^2+18M_{\eta}^4) \nonumber\\
&& -36M_{\pi}^2(11M_{\pi}^4-60M_{\pi}^2M_{\eta}^2+9M_{\eta}^4)B'(s_0,M_K^2,M_K^2)\nonumber\\
&&-12M_{\pi}^2(7M_{\pi}^2-3M_{\eta}^2)(15M_{\pi}^4-20M_{\pi}^2M_{\eta}^2-3M_{\eta}^4)B''(s_0,M_K^2,M_K^2) \nonumber\\
&&+M_{\pi}^2(M_{\eta}^2-M_{\pi}^2)(7M_{\pi}^2-3M_{\eta}^2)(3M_{\pi}^2+M_{\eta}^2)(27M_{\pi}^2+M_{\eta}^2)B^{(3)}(s_0,M_K^2,M_K^2)\nonumber\\
&&-2M_{\pi}^2(M_{\eta}^2-M_{\pi}^2)^2(3M_{\pi}^2+M_{\eta}^2)^3B^{(4)}(s_0,M_K^2,M_K^2) \nonumber\\
&& -36(M_{\pi}^2+3M_{\eta}^2)^2(2M_{\pi}^2+3M_{\eta}^2)B'(s_0,M_{\pi}^2,M_{\pi}^2)
+12(M_{\pi}^2+3M_{\eta}^2)^2(3M_{\pi}^4-4M_{\pi}^2M_{\eta}^2-3M_{\eta}^4)B''(s_0,M_{\pi}^2,M_{\pi}^2) \nonumber\\
&& -2(M_{\eta}^2-M_{\pi}^2)(M_{\pi}^2+3M_{\eta}^2)^2(3M_{\pi}^2+M_{\eta}^2)B^{(3)}(s_0,M_{\pi}^2,M_{\pi}^2)\,\}\,, \\
\tilde{\alpha}_{\mathrm{em}}^{(1)} &=& \frac{1}{54F_{\pi}^2}\dfrac{M_{\eta}^2(M_{\eta}-3M_{\pi})^2}{M_{\pi}^2+3M_{\eta}^2}\,\mathrm{Re}\,\{ 27(4\pi )^{-2}+72M_{\pi}^2B'(s_0,M_K^2,M_K^2)
+3(7M_{\pi}^2-3M_{\eta}^2)(3M_{\pi}^2+M_{\eta}^2)B''(s_0,M_K^2,M_K^2) \nonumber\\
&& -(M_{\eta}^2-M_{\pi}^2)(3M_{\pi}^2+M_{\eta}^2)^2B^{(3)}(s_0,M_K^2,M_K^2)\,\}\,,
\end{eqnarray}
\end{widetext}
expressed in terms of one- and two-point functions. 
In dimensional regularization, these take the following form
\begin{widetext}
\begin{eqnarray}
 A(m^2) &=& m^2\left[ -2\bar{\lambda}-\dfrac{1}{16\pi^2}\ln\left( \dfrac{m^2}{\mu^2}\right) \right] \,, \nonumber\\
 B(p^2,m_1^2,m_2^2)& =& -2\bar{\lambda}-\dfrac{1}{16\pi^2}(1-\ln\mu^2)
  -\dfrac{1}{16\pi^2}\int_0^1dx\ln [xm_1^2+(1-x)m_2^2-x(1-x)p^2]\,.
\end{eqnarray}
 \end{widetext}
The symbol $\mu$ denotes an ultraviolet scale contained in
\begin{equation}
 \bar{\lambda}=-\dfrac{1}{32\pi^2}\left[ \dfrac{2}{4-D}+\ln (4\pi )+1-\gamma_{\mathrm{E}}\right]\,,
\end{equation}
where $\gamma_{\mathrm{E}}$ is Euler's constant and $D$ the spacetime dimension.

\begin{acknowledgments}
A. Deandrea wishes to thank P. Santorelli for discussion on the $\eta \to 3 \pi$ in the early stage of this work.
This work was partially supported by European Commission RTN Programs MRTN-CT-2004-005104 and MRTN-CT-2006-035482.
\end{acknowledgments}

\end{document}